\documentclass[conference]{IEEEtran}
\IEEEoverridecommandlockouts

\usepackage{amsmath,amssymb,amsfonts}
\usepackage{graphicx}
\usepackage{breqn}
\usepackage{textcomp}
\usepackage{amsthm, bm, bbm}

\DeclareMathOperator*{\argmax}{arg\,max}

\DeclareMathOperator\atanh{atanh}
\usepackage[font=small,labelfont=bf]{caption}
\usepackage{pdfpages, float}
\usepackage{mathrsfs}
\usepackage{graphicx}

\newtheorem{assumption}{Assumption}

\newtheorem{definition}{Definition}

\newtheorem{example}{Example}

\usepackage{subcaption}
\definecolor{blu}{RGB}{0, 102, 204}
\definecolor{purp}{RGB}{128,0,128}
\definecolor{rd}{RGB}{255,69,0}
\definecolor{org}{RGB}{255, 95, 31}
\definecolor{cyn}{RGB}{0, 200, 200}

\usepackage[linesnumbered,ruled,vlined]{algorithm2e}
\usepackage{algpseudocode}
\usepackage{setspace}

\SetCommentSty{mycommfont}

\date{}
\begin{document}

\title{
	\vspace{-1cm}
	Reinforcement Learning for Sequential Decoding \\
	of Generalized LDPC Codes 
	\thanks{{The material is based upon work supported by the National Science Foundation under Grant Nos. CCF-2145917 and CNS-2148358.}}
}

\vspace{-1cm}
\begin{tiny}
	\author{
	\IEEEauthorblockN{
	Salman Habib and David G. M. Mitchell
		}
	\IEEEauthorblockA{
		Klipsch School of Electrical and Computer Engineering, New Mexico State University, Las Cruces, NM, 88003 \\
	{Email: \{salmanh, dgmm\}@nmsu.edu}
	}
}
\end{tiny}

\vspace{-1.5cm}
\maketitle

\begin{abstract} 
	In this work, we propose reinforcement learning (RL) for sequential decoding of moderate length generalized low-density parity-check (GLDPC) codes. Here, sequential decoding refers to scheduling all the generalized constraint nodes (GCNs) and single parity{-check nodes} (SPCNs) of a GLDPC code serially in each iteration. A GLDPC decoding environment is modeled as a finite Markov decision process (MDP) in which the state-space comprises of all possible sequences of hard-{decision} values of the variables nodes (VNs) connected to the scheduled GCN or SPCN, and the action-space of the MDP consists of all possible actions (GCN and SPCN scheduling). The goal of RL is to determine an optimized scheduling policy, {\emph{i.e.,}} one that results in a decoded codeword by minimizing the {complexity} of the belief propagation (BP) decoder. For training, we consider the proportion of correct bits at the output of the GCN or SPCN as a reward once it is scheduled. The expected rewards for scheduling all the GCNs/SPCNs in the code's Tanner graph are earned via BP decoding during the RL phase. The proposed RL-based decoding scheme {is shown to } significantly outperform the standard BP {flooding decoder, as well as a} sequential decoder in which the GCNs/SPCNs are scheduled randomly. 
\end{abstract}

\section{Introduction}
\label{sec:Intro}

Short and low rate ($R<1/2$) forward error correction schemes are critical for ultra reliable low latency communications (URLLC) in the {fifth generation (5G)} communications and beyond \cite{LOM18}, {\cite{Liu20}}. However, the design of such codes for attaining both low error-floor and good decoding threshold is a challenging problem \cite{Mao01}. Generalized low-density-parity-check (GLDPC) codes are one class of promising candidates which are constructed by replacing one or more single parity{-check} nodes (SPCNs) of {an} LDPC code {Tanner graph} by generalized constraint nodes (GCNs), which have multiple parity constraints \cite{Tan81}, {\cite{Olmos}}. The resulting GLDPC code has the same length as the base {(non-generalized)} code, but possess{es} a rate that decreases with increasing proportion of GCNs nodes in the Tanner graph of the code. A GCN represents the parity-check constraints induced by a separate component code, such as a $[p,k]$ Hamming code, where both the component code length $p$, and number of information bits $k$, are typically small. Replacing a SPCN with a GCN improves the node's decoding capability at the expense of the {associated} rate loss. The component codes can be decoded via trellis based decoders such as the Bahl-Cocke-Jelinek-Raviv (BCJR) algorithm. However, this scheme is computationally intensive and may be less attractive for applications which are power constrained. An alternative method is to decode the GCN/SPCN subgraph via the belief propagation (BP) algorithm which has a significantly lower decoding complexity. 

The decoding performance of both these schemes for quasi-cyclic GLDPC codes {with Hamming component codes} have been analyzed in \cite{Liva10}. The quasi-cyclic structure is a property of the base {(non-generalized) Tanner graph that is desirable for efficient implementation.} By carefully selecting the quasi-cyclic base graph, the authors were able to construct GLDPC codes with excellent waterfall and error-floor performances. Recently, GLDPC codes have also been proposed for URLLC \cite{LOM18}. The authors study the optimal proportion of GCNs of multiple GLDPC codes constructed from $(\gamma,p)$-regular LDPC base codes, where $\gamma$ and $p$ represent the column and row weights of the code, respectively. Here, optimality refers to minimizing the GLDPC code's asymptotic gap to capacity, which is an important metric for gauging the finite-length behavior of iteratively decodable codes. Specifically, they consider $\gamma=2$ and $p=6, 7,$ and $15$. It is shown that the optimal proportion, $\mu$, of GCN nodes is $0.75$ (resp., $0.8$) in case of $(2,6)$-regular and $(2,7)$-regular (resp., $(2,15)$-regular) codes. These results were obtained by randomly replacing the SPCNs of the base code with {Hamming code} GCNs. 


\indent In this work, we propose reinforcement learning (RL) for {sequential} decoding {of} GLDPC codes{. Related work includes an RL aided bit-flipping decoder proposed in \cite{CHMRP19}, which was shown to outperform a standard bit-flipping scheme for Reed-Muller and BCH codes. T}he performance of RL-based sequential soft-decoding of LDPC codes {was} examined in \cite{JSAIT,reldec}. It was shown that RL is able to learn an optimized {SPCN} scheduling order, {\emph{i.e.,}} one which results in a codeword output by minimizing the expected number of {SPCN} to VN message propagations for a given LDPC {Tanner graph}. 

{Here, we consider sequential decoding of GLDPC codes with double BP decoding, \emph{i.e.,} BP decoding of the base Tanner graph and of the component codes, which are either fully generalized (all SPCNs of the base LDPC code are replaced by Hamming component codes), or partially generalized (a certain proportion, $\mu$, of the SPCNs are replaced with GCNs). It is shown that a sequential decoding schedule can improve the {frame error rate (FER)} performance of GLDPC codes, as well as reducing the convergence speed and associated complexity, measured as the number of constraint node computations. This is particularly important for decoding GLDPC codes, where the decoder complexity is dominated by GCN computation \cite{LOM18}. In particular, we show that the RL optimized scheme can improve the performance of a moderate length ($469$ bits) GLDPC code by approximately $1$ dB at a target FER of $10^{-4}$ with complexity saving of greater than $50$\% at high signal-to-noise ratios (SNRs).}

\section{Preliminaries}

\subsection{Generalized Low-density Parity-check Codes}

Let $\mathbf{H}_B\in \mathbb{F}_2^{m\times n}$ be the parity-check matrix of a $(\gamma,p)${-regular} base LDPC code with {(base)} Tanner graph $G_B$, and let $\mathbf{H}_C\in \mathbb{F}_2^{(p-k)\times p}$ be the parity-check matrix of a $[p,k]$ component  code. By $\mu=g/m, g\in \{0,\ldots,m\}$, we denote the fraction of  SPCNs in $G_B$ that are replaced with GCNs, where the resulting graph, referred to as the generalized Tanner graph $G$, contains $g$ GCNs and $m-g$ SPCNs. Note that for $g=0$, $G$ is identical to $G_B$. Suppose $\zeta$ is the set of SPCN indices of $G_B$ that are replaced with GCNs, and let $\zeta'=\{0,.\ldots,m-1\}\setminus \zeta$ be the set of remaining SPCN indices in $G_B$. A GLDPC code with parity-check matrix $\mathbf{H}\in \mathbb{F}_2^{(m+\mu m(p-k-1))\times n}$ and rate $R=(n-\text{rank}(\mathbf{H}))/n$ is obtained by replacing the non-zero (resp., zero) entries in the rows of $\mathbf{H}_B$, corresponding to SPCNs in $\zeta$, with a column of $\mathbf{H}_C$ (resp., all-zero column of length $p-k$){, where} all rows of $\mathbf{H}$ representing the SPCNs are identical to the rows of $\mathbf{H}_B$ that correspond to the SPCNs in $\zeta'$. Note that due to this construction, each of the GCNs/SPCNs in $G$ are connected to $p$ VNs. {Note also that we do not decode with the parity-check matrix $\mathbf{H}$. Rather, we decode with BP decoding on the generalized Tanner graph $G$, where GCNs are decoded according to that particular linear code via component BP decoding. Since a constraint node of a GLDPC code can be either a GCN or an SPCN, we call it simply a constraint node (CN) in the rest of the paper.}

\subsection{Reinforcement Learning}
In an RL problem, an agent (learner) interacts with an environment whose \textit{state space} can be modeled as a finite Markov decision process (MDP) \cite{Sutton18}. The agent takes \textit{actions} that {alter} the state of the environment and receives a \textit{reward} in return for each action, with the goal of maximizing the total reward in a series of actions. The optimized sequence of actions is obtained by employing a policy which utilizes an \textit{action-value function} to determine how beneficial an action is for maximizing the long-term expected reward. Suppose that an environment allows $m$ possible actions, and let {the random variable} ${A^{(\ell)}}\in \{0,\ldots,m-1\}$ with realization $a$ represent the index (a scheduled {CN} {update} for example) of an action take by the agent during learning step $\ell\in \{0,\ldots, \ell_\max-1\}$. Let ${S^{(\ell)}}$ with realization $s^{(\ell)}\in \mathbb{Z}$, represent the current state of the environment before taking action ${A^{(\ell)}}$ and let ${S^{(\ell+1)}}$ with realization ${s^{(\ell+1)}}$ represent a new state of the MDP after executing ${A^{(\ell)}}$. Let a state space $\mathcal{S}$ contain all possible state realizations. Also, let $R_{\ell}=R({S^{(\ell)}},{A^{(\ell)}},{S^{(\ell+1)}})$ be the reward yielded at step $\ell$ after taking action ${A^{(\ell)}}$ in state ${S^{(\ell)}}$ which will yield state ${S^{(\ell+1)}}$. Optimal policies for MDPs can be estimated via Monte Carlo techniques, such as Q-learning. The estimated action-value function $Q_\ell({S^{(\ell)}},{A^{(\ell)}})$, also known as the Q-function, represents the expected long-term reward an agent obtains after taking action ${A^{(\ell)}}$ in state ${S^{(\ell)}}$. Learning involves iteratively adjusting the action-value function for a specific  $({S^{(\ell)}},{A^{(\ell)}})$ pair, based on  previously learned action-values for the same state and action pair and the reward $R_\ell$ earned from taking that action. Learning results in an optimal policy that guides the agent to select an action in a given state that maximizes the Q-function value for that state and action pair during inference.

\subsection{Learning Environment for GLDPC Decoding}

Let $\mathbf{x}=[x_0,\ldots,x_{n-1}]$ and $\mathbf{y}=[y_0,\ldots,y_{n-1}]$ represent the transmitted and the received words respectively where for each $v\in \{0,\ldots,n-1\}$, $x_v\in \{0,1\}$ and $y_v=(-1)^{x_v}+z$ with $z\sim \mathcal{N}(0,\sigma^2)$. The posterior log-likelihood ratio (LLR) of $x_v$ is expressed as $L_v=\log \frac{\Pr(x_v=0|y_v)}{\Pr(x_v=1|y_v)}$. Let ${\hat{L}_v^{(\ell)}}=\sum_{c\in \mathcal{N}(v)} m_{c\rightarrow v}^{(\ell)}+L_v$ be the posterior LLR computed by VN $v$ during iteration $\ell$, where $\mathcal{N}(v)$ denotes the set of neighboring SPCNs of VN $v$, ${\hat{L}_v^{(0)}}=L_v$,  and $m_{c\rightarrow v}^{(\ell)}$ is the message received by VN $v$ from neighboring {SPCN} $c$ in iteration $\ell$ computed based on standard BP as 
\begin{equation}
	m_{c\rightarrow v}^{(\ell)}=2\atanh \prod_{v'\in \mathcal{N}(c)\setminus v} \tanh\left(\frac{m_{v'\rightarrow c}^{(\ell-1)}}{2}\right).
	\label{eq:mcv}
\end{equation}
Here, $\mathcal{N}(c)$ denotes the set of neighboring VNs of $c$, and
\begin{equation}
	m_{v\rightarrow c}^{(\ell)}= \sum_{c'\in \mathcal{N}(v)\setminus c} m_{c'\rightarrow v}^{(\ell)}+L_v
	\label{eq:mvc}
\end{equation}
is the message propagated from VN $v$ to {SPCN} $c$. Moreover, let ${\hat{L}_{j,a}^{(\ell)}}$ be the posterior LLR computed during learning step $\ell$ by VN $j$ in the subgraph induced by a {CN} with index $a\in\{0,\ldots,m-1\}$. Hence, ${\hat{L}_{j,a}^{(\ell)}=\hat{L}_v^{(\ell)}}$ if VN $v$ in the Tanner graph is also the $j$-th VN in the subgraph induced by the {CN} with index $a$. Let $\mathbf{\hat{x}}_a^{(\ell)}=[\hat{x}_{0,a}^{(\ell)},\ldots,\hat{x}_{l_a-1,a}^{(\ell)}]$, where $l_a$ is the number of VNs adjacent to {CN} with index $a$. After scheduling {CN} $a$, the state of the MDP associated with this node is given by its output $\mathbf{\hat{x}}_a^{(\ell)}$ that is obtained by taking hard decisions on the vector of posterior LLRs {$\mathbf{\hat{L}}_a^{(\ell)}=[\hat{L}_{0,a}^{(\ell)}, \ldots, \hat{L}_{l_a-1,a}^{(\ell)}]$}, computed according to 
\begin{equation}
	\hat{x}_{j,a}^{(\ell)}=
	\begin{cases}
		0, \text{ if } {\hat{L}_{j,a}^{(\ell)}} \geq 0,\\
		1, \text{ otherwise.}
	\end{cases}
	\label{eq:hd}
\end{equation}
We assign a unique index $s_a^{(\ell)}\in \{0,\ldots,2^{l_a}-1\}$ to a realization of $\mathbf{\hat{x}}_a^{(\ell)}$, and thus we call $s_a^{(\ell)}$ the state of our MDP in the remainder of the paper. Since the state of the MDP is a sequence of hard-decision VN values associated with the subgraph of the scheduled node, the collection of signals $\mathbf{\hat{x}}_0^{(\ell)},\ldots,\mathbf{\hat{x}}_{m-1}^{(\ell)}$ at the end of decoder iteration $\ell$ forms the entire state of the MDP associated with our learning scheme. 
 
Given that the transmitted signal $\mathbf{x}$ is known during the training phase, let $\mathbf{x}_a=[x_{0,a},\ldots,x_{l_a-1,a}]$ be a vector containing the $l_a$ bits of $\mathbf{x}$ that are reconstructed as $\mathbf{\hat{x}}_a^{(\ell)}$. In each learning step $\ell$, the reward $R_a$ obtained by the agent after scheduling {CN} $a$ is defined as 
\begin{equation}
	\label{eq:rew}
	R_a=\frac{1}{l_a} \sum_{j=0}^{l_a-1} \mathbbm{1}(x_{j,a} = \hat{x}_{j,a}),
\end{equation}
where $\mathbbm{1(}\cdot)$ denotes the indicator function. Thus, the reward earned by the agent after scheduling {CN} $a$ is identical to the probability that the transmitted bits $x_{0,a},\ldots,x_{l_a-1,a}$ are correctly reconstructed.

\section{RL for Sequential GLDPC Decoding}
Our goal is to learn an  optimized sequence of actions, \emph{i.e.}, the scheduling of individual {CNs} $\mathcal{C}_1,\ldots,\mathcal{C}_{m}$ {where} $\mathcal{C}_i=\{c_{i,1},\ldots,c_{i,z}\}$, $c_{i,j}\in \{0,\ldots,m-1\}$ is the index of the $j$-th SPCN in the $i$-th CN, {and} $z$ is the number of SPCNs involved in the {CN} (which is $1$ in case of an SPCN). By $\mathcal{N}(\mathcal{C}_i)$ we refer to all the VNs connected to the $i$-th {CN}. We choose the BP decoding algorithm {to learn} the optimized {CN} scheduling order. We form the subgraph of a CN, which consists of the target CN, all attached VNs, as well as their attached CNs. {Then a} single {CN} scheduling step involves CN computation, message propagation from {the} SPCN to {its} neighboring VNs in the first half of the {step}, and message propagation from these VNs to their {SPCN} neighbors in the second half.\footnote{Note that any component decoder could be used to compute the {SPCN} to VN messages.} In other words, a {CN} subgraph is decoded via flooding{, where in this paper, the {SPCN} to VN messages are computed with a BP decoder and flooding schedule on the component graph.} Every {CN} is scheduled exactly once within a single decoder iteration. Sequential {CN} scheduling is carried out until a stopping condition is reached, or an iteration threshold is exceeded. 

\begin{figure}[h]
  \centering
  \includegraphics[scale=0.5]{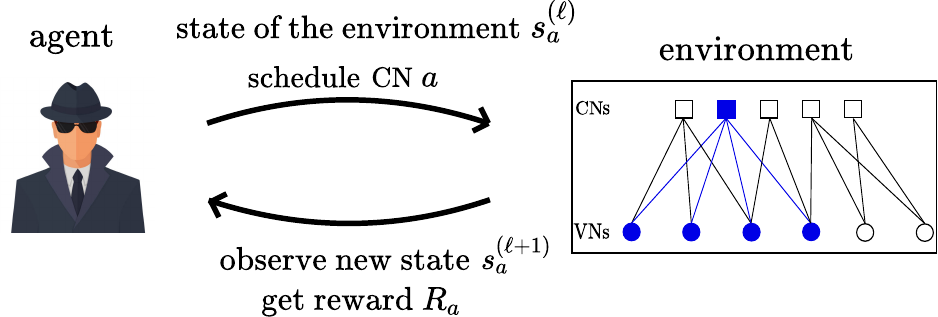}
  \caption{Illustration of {our} GLDPC scheduling policy learning framework. In each learning step, a fictitious agent schedules a {CN} with index $a$ when the environment state, based on hard-decisioned (blue) VN values, is $s_a^{(\ell)}$. Once an action is taken, the state of the environment changes from $s_a^{(\ell)}$ to ${s_a^{(\ell+1)}}$, as the blue VN values are updated after scheduling, and the agent receives reward $R_a$ that indicates the accuracy of the hard-decisions taken by the BP algorithm for each blue VN.}
  \label{fig:RL}
\end{figure}

The MDP for sequential decoding is shown in Fig. \ref{fig:RL}. The idea is that scheduling a {CN} with index $a$, represented by the blue square, alters the hard-decisioned beliefs of the blue VNs associated with the node, \emph{i.e.,} the state of the environment changes from $s_a^{(\ell)}$ to ${s_a^{(\ell+1)}}$. In return, the agent receives a reward $0\leq R_a\leq 1$ according to (\ref{eq:rew}) which grows in proportion to the number of correct bits of output $\mathbf{\hat{x}}_a^{(\ell)}$. 
The long-term expected rewards, for scheduling a {CN} with index $a$ in state $s_a^{(\ell)}$, is learned via standard Q-learning \cite{Watkins89}. 

{Now, we consider learning the CN scheduling policy in two distinct ways: in the first approach, we learn a single policy using a dataset of LLR vectors corresponding to a range of $K$ SNR values, and in the second, separate policies are learned using a specific dataset for each fixed SNR.} Let an event $M$ be true if {the first approach is employed, and false otherwise.} We then define a learning indicator function

\begin{equation}
	\label{eq:egreedy}
	\mathbb{I}_M\triangleq
	\begin{cases}
	 1, \text{ if event } M \text{ is true,} \\
     0, \text{ otherwise}. 
	\end{cases}
\end{equation}

During the learning phase, the long-term expected rewards are earned according to the recursion
\begin{align}
	\begin{split}
	G(\ell+1,s_a^{(\ell)},&a,\mathbb{I}_M) = (1-\alpha)G(\ell,s_a^{(\ell)},a,\mathbb{I}_M) + \\ &\alpha \Bigl(R_a+\beta \max_{a'\in \{0,\ldots,m-1\}}G(\ell,{s_a^{(\ell+1)}},a',\mathbb{I}_M) \Bigr),
	\label{eq:q_cls2}
	\end{split}
\end{align}
\noindent where
\vspace{-0.5cm}
\begin{equation}
	\label{eq:egreedy}
	G({\ell},s_a^{(\ell)},a,\mathbb{I}_M)=
	\begin{cases}
	 Q_{{\ell}}(s_a^{(\ell)},a), \text{ if }\mathbb{I}_M=1, \\
    Q_{{\ell}}^{(k)}(s_a^{(\ell)},a), \text{ otherwise}, 
	\end{cases}
\end{equation}
${s_a^{(\ell+1)}}$ represents the new state of the MDP after taking action $a$ in state $s_a^{(\ell)}$, $0<\alpha <1$ is the \textit{learning rate}, $0<\beta<1$ is the \textit{reward discount rate}, and $\ell$ is the number of learning steps elapsed after observing the initial state $s_a^{(0)}$ corresponding to a received channel output $\mathbf{L}=[L_0,\ldots,L_{n-1}]$ in a learning episode{. A learning episode comprises all the Q-learning steps needed to learn the action-values corresponding to a single training example (LLR vector).} Here, note that {$G({\ell+1},s_a^{(\ell)},a,1)$} is a future action-value resulting from action $a$ in the current state $s_a^{(\ell)}$ when learning is done over a mixture of SNR values chosen from a set $\mathcal{S}=\{S_1,\ldots,S_K\}$, where $S_i\in \mathbb{R}$ is the $i$-th SNR value, and $K$ is the total number of distinct SNRs considered for training (a single policy is learned for each SNR). In contrast, {$G({\ell+1},s_a^{(\ell)},a,0)$} is a future action-value when Q-learning is done only for the $k$-th SNR value in $\mathcal{S}$ (a separate policy is learned for each SNR). For each $\ell$, {CN} $a$  is selected via an $\epsilon$-greedy approach according to
 \begin{equation}
	\label{eq:egreedy}
	a=
	\begin{cases}
     \text{selected uniformly at random w.p. } \epsilon \text{ from }\mathcal{A}, \\
	 \pi(s_a^{(\ell)},\mathbb{I}_M) \text{ selected w.p. } 1-\epsilon, 
	\end{cases}
\end{equation}
where $\epsilon$ is the probability of exploration, $\mathcal{A}=\{0,\ldots,m\}$ is a set of all possible actions, and $\pi(s_a^{(\ell)},\mathbb{I}_M)$ is an agent's  policy for taking an action in state $s_a^{(\ell)}$, expressed as
\begin{equation}
	\label{eq:pol_clus}
	\pi(s_a^{(\ell)},\mathbb{I}_M)=\argmax_{a \in \{0,\ldots,m-1\}} G(\ell,s_a^{(\ell)},a,\mathbb{I}_M).
\end{equation}
At the end of the learning phase, we obtain an optimal {CN} scheduling policy that maximizes the long-term expected reward in state $s_a^{(\ell)}$, given by 
\begin{equation}
	\label{eq:pi_Qopt}
	\pi^*(s_a^{(\ell)},\mathbb{I}_M)=\argmax_{a} G^*(\ell,s_a^{(\ell)},a,\mathbb{I}_M),
\end{equation}
where $G^*(\ell,s_a^{(\ell)},a,\mathbb{I}_M)$ is the highest possible long-term expected reward (action-value) for a given $(s_a^{(\ell)},a)$ pair. Since $G^*(\ell,s_a^{(\ell)},a,\mathbb{I}_M)$ is unknown, the outcome of Q-learning is an \emph{optimized} {CN} scheduling policy, $\hat{\pi}(s_{a_i}^{(I)},\mathbb{I}_M)$, for scheduling the $i$-th {CN} expressed as 
\begin{equation}
	\label{eq:pi_i2}
 \hat{\pi}(s_{a_i}^{(I)},\mathbb{I}_M)=\argmax_{a_i\in \{0,\ldots,m-1\}\setminus \{a_0,\ldots, a_{i-1}\}} g(s_{a_i}^{(I)},a_i,\mathbb{I}_M), 
\end{equation} 
\noindent where 
 \begin{equation}
	\label{eq:gsa}
	g(s_{a_i}^{(I)},a_i,\mathbb{I}_M)=
	\begin{cases}
	 \hat{Q}(s_{a_i}^{(I)},a_i), \text{ if }\mathbb{I}_M=1, \\
    \hat{Q}^{(k)}(s_{a_i}^{(I)},a_i), \text{ otherwise}, 
	\end{cases}
\end{equation}
$i\in \{0,\ldots,m-1\}$, $I$ is the decoder iteration during inference, and $a_i$ indicates the {CN} index to be scheduled at time instant $i$. Further, $\hat{Q}(s_{a_i}^{(I)},{a_i})$ and $\hat{Q}^{(k)}(s_{a_i}^{(I)},a_i)$ represent optimized action-value functions once learning ends. {Note that we draw distinction between the decoder iteration $\ell$ during RL and the decoder iteration $I$ during inference, as the maximum number of iterations, $I_\max$, for the latter need not match $\ell_\max$.}  

The proposed learning scheme is shown in Algorithm \ref{alg:Qlrn}. The input to this algorithm is the learning indicator function $\mathbb{I}_M$, the code's parity-check matrix $\mathbf{H}$, and a set $\mathscr{\hat{L}}$ (resp., $\mathscr{L}_k$) of $|\mathscr{\hat{L}}|$ (resp., $|\mathscr{L}_k|$) realizations of $\mathbf{L}$ over which standard Q-learning is performed. Note that each vector in $\mathscr{\hat{L}}$ correspond{s} to an SNR selected from $\mathcal{S}$. That is, there are $|\mathscr{\hat{L}}|/K$ LLR vectors in $\mathscr{\hat{L}}$ with the same SNR. In comparison, all the LLR vectors in $\mathscr{L}_k$ correspond to a fixed SNR. Note also that the action-values $Q_0(s_a^{(0)},a)$ and $Q_0^{(k)}(s_a^{(0)},a), \text{ } \forall (s_a^{(0)},a),$ are initialized with zeros. Finally, we note that the {posterior} LLR vector $\mathbf{\hat{L}}_{I}=[\hat{L}_I^{(0)},\ldots,\hat{L}_I^{(n-1)}]$, used for making hard-decisions in Step 11 is initialized using $\mathbf{L}$ in Step 4. 

In Algorithm \ref{alg:Qlrn}, for a training sample $\mathbf{L} \in |\mathscr{L}|$, the initial state of all the {CNs} are determined in Step 5. Then, a total of $\ell_{\max}$ training rounds are completed in Steps 6-18. In each episode, an action is taken according to (\ref{eq:egreedy}) and a single flooding BP iteration is performed in the subgraph of the scheduled {CN} as shown in Steps 8-9. The new {CN} output is determined in Steps 10-12, leading to a new {CN} state in Step 13. The reward for taking action $a$ in Step 7 is earned in Step 14. Finally, the action-value (long-term reward) is computed in Step 15. The output of Algorithm \ref{alg:Qlrn} is an optimized {CN} scheduling policy $\hat{\pi}(s_{a_i}^{(I)},\mathbb{I}_M)$. {For inference, we invoke a BP decoding algorithm with inputs comprising of LLR vector $\mathbf{L}$ that corresponds to one of the SNRs in $\mathcal{S}$, parity-check matrix $\mathbf{H}$ of the GLDPC code, and $\mathbb{I}_M$. In every iteration, we generate an optimized CN scheduling order using $\hat{\pi}(s_{a_i}^{(I)},\mathbb{I}_M)$.}

{\linespread{0.5}\selectfont  
\begin{algorithm}
\caption{GLDPC scheduling policy learner}
\SetAlgoLined
\DontPrintSemicolon
\SetKwInOut{Input}{Input}
\SetKwInOut{Output}{Output}
\Input{learning indicator $\mathbb{I}_M$, training sets $\mathscr{\hat{L}}, \mathscr{L}_k$, parity-check matrix $\mathbf{H}$}
\Output{optimized scheduling policy $\hat{\pi}(s_{a_i}^{(I)},\mathbb{I}_M)$} 
\label{alg:Qlrn}

\textbf{Initialization:} For $\mathbb{I}_M=1$, initialize $\mathscr{L}\leftarrow \mathscr{\hat{L}}$ and $Q_0(s_a^{(0)},a)\leftarrow 0 \text{ } \forall (s_a^{(0)},a)$, otherwise initialize $\mathscr{L}\leftarrow \mathscr{L}_k$ and $Q_{0}^{(k)}(s_a^{(0)},a) \leftarrow 0 \text{ } \forall (s_a^{(0)},a)$\;


\For{each $\mathbf{L}\in \mathscr{L}$} { 
	$\ell \leftarrow 0$\;
	{$\mathbf{\hat{L}}_\ell\leftarrow \mathbf{L}$}\;
	{determine initial states of all GCNs using (\ref{eq:hd})}\;
	\tcp{start of an episode}
	\While{$\ell<\ell_{\max}$}
	{
	select {CN} $a$ according to {(\ref{eq:egreedy})}\;
	for each {SPCN} $c$ in $\mathcal{C}_a$, compute and propagate $m_{c\rightarrow v}^{(\ell)}$ $\forall v\in \mathcal{N}(c)$\;
	for each VN $v$ in $\mathcal{N}(\mathcal{C}_a)$, compute and propagate $m_{v\rightarrow c}^{(\ell)}$ $\forall c\in \mathcal{N}(v)$\;
	\tcp{determine CN output}
		\For{{each} VN $v$ in the subgraph of {CN} $a$}{
				If ${\hat{L}_v^{(\ell)}}\geq 0$, $\hat{x}_{v,a}^{(\ell)}\leftarrow 0$, otherwise $\hat{x}_{v,a}^{(\ell)}\leftarrow 1$\; 
}
		determine index ${s_a^{(\ell+1)}}$ of $\mathbf{\hat{x}}_a$ via binary to decimal conversion\;
		update $R_a$ according to (\ref{eq:rew})\;
		compute $G(\ell+1,s_a^{(\ell)},a,\mathbb{I}_M)$ according to (\ref{eq:q_cls2})\;
		$s_a^{(\ell+1)}\leftarrow {s_a^{(\ell+1)}}$\;
		$\ell\leftarrow \ell+1$\;
	}
}
\end{algorithm}
}

%
%
%
%
%
%

\section{Numerical Results}
\label{sec:results}
We compare the RL-based sequential BP decoding performance of GLDPC codes against both flooding and random sequential BP decoding schemes. The flooding decoder schedules all the GCNs and SPCNs serially, and then updates all the VNs in the Tanner graph simultaneously. In case of random sequential decoding, the GCNs and SPCNs are randomly scheduled in sequence. Once a {CN} is scheduled, only the VNs in its subgraph are updated at once, and this process repeats for all the {CNs}. Consequently, in the case of sequential decoding, a VN that is common to multiple {CN} subgraphs gets updated more than once in the same iteration. In contrast, each VN is updated only once per iteration in case of flooding.  

We utilize flooding, random sequential, and RL-based schemes for decoding GLDPC codes using a $(2,7)$ base LDPC code of length $n=469$, where the number of SPCNs is $m=134$. Multiple GLDPC codes are constructed for different values of $\mu$, which is the fraction of GCNs in the GLDPC code Tanner graph. A $[7,4]$ Hamming component code is used in each case. Specifically, we consider $\mu$ values of $0,0.373,0.746,0.970$, and $1$, resulting in GLDPC codes $\mathcal{C}_1,\mathcal{C}_2,\mathcal{C}_3,\mathcal{C}_4$, and $\mathcal{C}_5$, with rates $0.714, 0.501, 0.288,$ {$0.160,$} and $0.143$, respectively. The choice of GLDPC code length is influenced by the run-time of the proposed learning schemes on our system. We employ the GLDPC scheduling policy learner of Algorithm \ref{alg:Qlrn} using Q-learning. For both learning schemes, we found a reasonable choice of hyper-parameters that render superior decoding performance when compared to flooding and random sequential schemes. Those parameters include a learning rate of $\alpha=0.1$, a reward discount rate of $\beta=0.9$, a probability of exploration of $\epsilon=0.6$, and a maximum number of steps per learning episode of $\ell_{\max}=50$.

For learning we select $K=6$ different $E_b/N_0$ values (in dB) from the set $\mathcal{S}=\{1,2,3,4,4.5,5\}$. For $\mathbb{I}_M=1$, the GLDPC scheduling policy is learned for dataset size $|\mathscr{\hat{L}}|=180000$, ensuring that $1/K$-th of the dataset contains LLR vectors of a fixed SNR. That is, a single {CN} scheduling policy is learned using dataset for all $K$ SNR values in $\mathcal{S}$, and $|\mathscr{\hat{L}}|/K=30000$ training samples are allotted for each. On the other hand, for $\mathbb{I}_M=0$, the GLDPC scheduling policy is learned over a dataset of size $|\mathscr{L}_k|=30000$. That is, a separate {CN} scheduling policy is learned using each $\mathscr{L}_k$ for the $k$-th SNR value in $\mathcal{S}$. The training set sizes $|\mathscr{\hat{L}}|$ and $|\mathscr{L}_k|$ are chosen to ensure that the dataset is large enough for accurate training without incurring too much simulation time. 

For both training and inference, we transmit all-zero codewords using BPSK modulation. Note that training with the all-zero codeword is sufficient since, due to the symmetry of the BP decoder and the channel, the decoding error is independent of the transmitted signal (see \cite[Lemma 4.92]{RU08}). For performance measures, {we consider the FER} given by $\Pr[\mathbf{\hat{x}} \neq \mathbf{x}]$ and choose a maximum number of BP iterations $I_\max=\ell_\max$. 

\begin{figure}[h]
  \centering
  \includegraphics[scale=0.45]{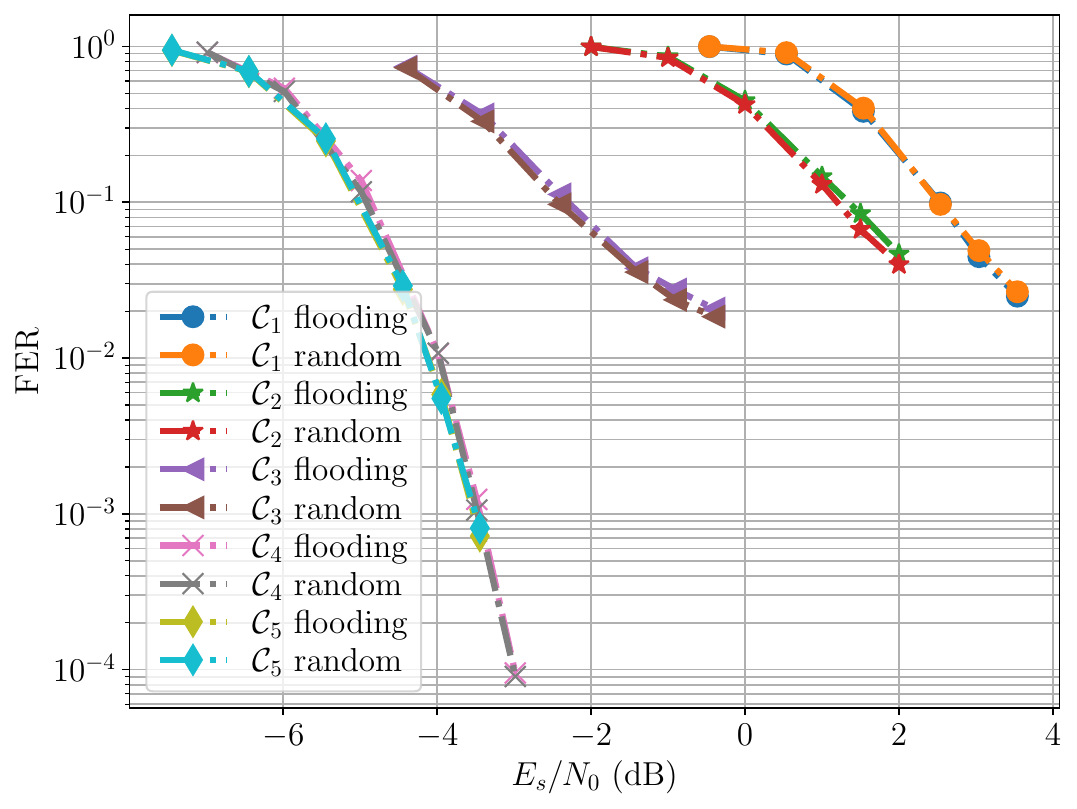}
  \vspace{-0.12cm}
  \caption{FER results using the flooding and random sequential decoding scheme for different GLDPC codes of block length $n=469$.}
  \label{fig:res2}
\end{figure}
\begin{figure}[h]
  \centering
  \includegraphics[scale=0.45]{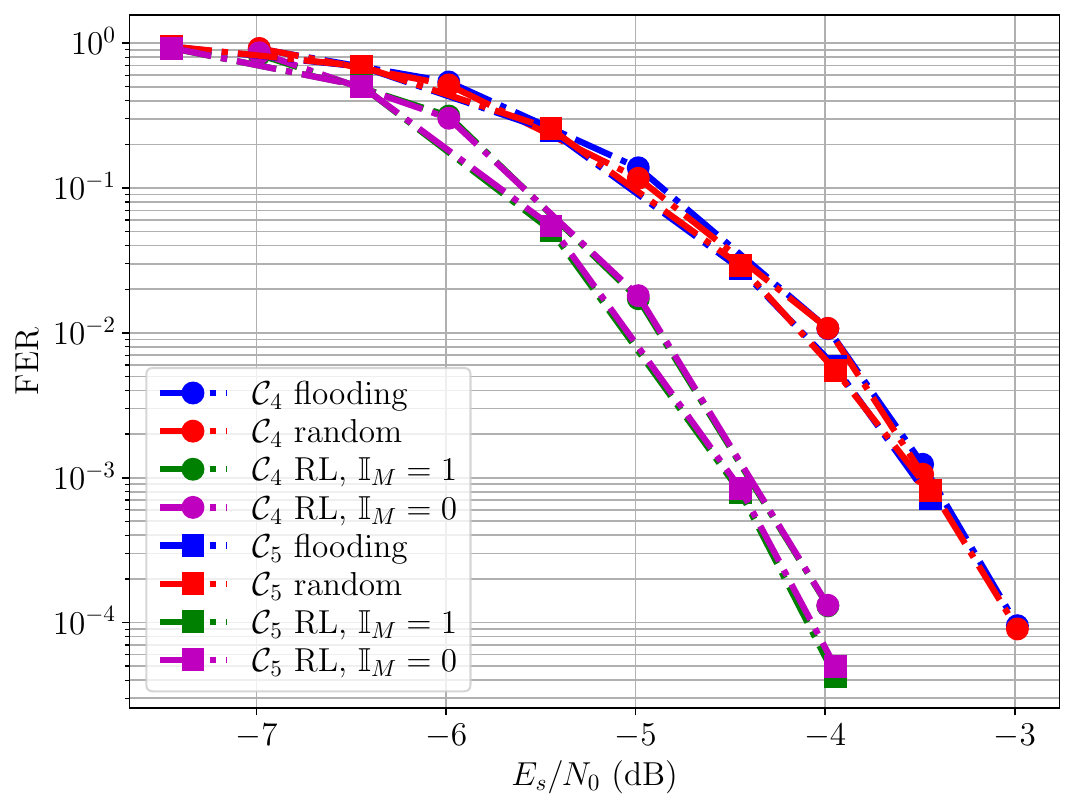}
  \vspace{-0.12cm}
  \caption{FER results using different BP decoding schemes for GLDPC codes $C_4$ and $C_5$.}
  \label{fig:res4}
\end{figure}

The curves FER vs. channel SNR, in terms of {$E_s/N_0$} in dB, for codes $\mathcal{C}_1,\ldots,\mathcal{C}_5$ using the flooding and random sequential decoding techniques are shown in Fig. \ref{fig:res2}. This figure reveals that, as expected, the GLDPC decoding performance generally improves as $\mu$ increases, and the random sequential decoder slightly outperforms flooding (note that we plot in terms of $E_s/N_0$). Furthermore, in Fig. \ref{fig:res4}, we clearly see significant gains of RL-based decoding in case of both $\mathbb{I}_M=1$ and $\mathbb{I}_M=0$, when compared to flooding and random sequential decoding schemes for codes $\mathcal{C}_4$ and $\mathcal{C}_5$. Similar gains are obtained for {$\mathcal{C}_2$ and $\mathcal{C}_3$ (not shown). For this example, RL-based sequential decoding does not provide any gain for $\mathcal{C}_1$.} We also note that the performances using the two RL schemes are almost identical. 

In Table \ref{tab:tab}, we compare the average number of {SPCN} to VN messages propagated in the considered decoding schemes to attain the results in Fig. \ref{fig:res4} for the $\mathcal{C}_5$ code. We note that the RL-based scheme (for $\mathbb{I}_M=1$) require a significantly lower number of {SPCN} to VN messages when compared to the other decoding schemes (the complexity results for $\mathbb{I}_M=0$ are similar to the $\mathbb{I}_M=1$ case, and hence omitted).

{\linespread{0.85}
\begin{table}[h]
\small
\centering
\resizebox{\columnwidth}{!}{
	\begin{tabular}{|c|c|c|c|c|c|}
		\hline
		$\mathbf{E_s/N_0}$ \textbf{(dB)}
		&$\mathbf{-7.45}$
		&$\mathbf{-6.45}$ 
		&$\mathbf{-5.45}$  
		&$\mathbf{-4.45}$  
		&$\mathbf{-3.94}$ \\
		
		
		\hline\hline
		flooding&$78924$&$68751$&$42087$&$21498$&$16217$\\
		\hline
		random seq.&$78342$&$65248$&$36664$&$15469$&$11170$\\
		\hline
		RL ($\mathbb{I}_M=1$)&$77640$&$56127$&$20561$&$9214$&$7381$\\
		\hline
	\end{tabular}
%
%
}
	\vspace{0.2cm}
	\caption{\label{tab:tab} Average number of {SPCN} to VN messages propagated by different decoders for $\mathcal{C}_5$ to attain the results shown in Fig. \ref{fig:res4}. }
\end{table}
}

\section{Conclusion}

We studied sequential (double) BP decoding  of generalized LDPC codes with the aid of RL. In learning, a scheduling instant updates all the nodes of a {CN} subgraph via flooding. By learning the {CN} scheduling order, we are able to outperform both a random {CN} scheduling scheme, as well as a flooding scheme, where the gain increases with the proportion of GCN nodes in the GLDPC Tanner graph. Moreover, the number of CN updates was greatly reduced, particularly at high SNRs. Future work will take into account other component code decoding schemes, such as the BCJR algorithm.


\bibliographystyle{IEEEtran}
\bibliography{Bib}

\begin{thebibliography}{10}
\providecommand{\url}[1]{#1}
\csname url@samestyle\endcsname
\providecommand{\newblock}{\relax}
\providecommand{\bibinfo}[2]{#2}
\providecommand{\BIBentrySTDinterwordspacing}{\spaceskip=0pt\relax}
\providecommand{\BIBentryALTinterwordstretchfactor}{4}
\providecommand{\BIBentryALTinterwordspacing}{\spaceskip=\fontdimen2\font plus
\BIBentryALTinterwordstretchfactor\fontdimen3\font minus
  \fontdimen4\font\relax}
\providecommand{\BIBforeignlanguage}[2]{{%
\expandafter\ifx\csname l@#1\endcsname\relax
\typeout{** WARNING: IEEEtran.bst: No hyphenation pattern has been}%
\typeout{** loaded for the language `#1'. Using the pattern for}%
\typeout{** the default language instead.}%
\else
\language=\csname l@#1\endcsname
\fi
#2}}
\providecommand{\BIBdecl}{\relax}
\BIBdecl

\bibitem{LOM18}
Y.~Liu, P.~M. Olmos, and D.~G.~M. Mitchell, ``Generalized {LDPC} codes for
  ultra reliable low latency communication in 5{G} and beyond,'' \emph{IEEE
  Access}, vol.~6, pp. 72\,002--72\,014, 2018.

\bibitem{Liu20}
------, ``On the design of generalized {LDPC} codes with component {BCJR}
  decoding,'' in \emph{GLOBECOM 2020 - 2020 IEEE Global Communications
  Conference}, 2020, pp. 1--6.

\bibitem{Mao01}
Y.~Mao and A.~Banihashemi, ``A heuristic search for good low-density
  parity-check codes at short block lengths,'' in \emph{ICC 2001. IEEE
  International Conference on Communications}, vol.~1, 2001, pp. 41--44 vol.1.

\bibitem{Tan81}
R.~M. Tanner, ``A recursive approach to low complexity codes,'' \emph{IEEE
  Trans. on Inf. Theory}, vol.~27, no.~5, pp. 547--553, Sep 1981.

\bibitem{Olmos}
P.~M. Olmos, D.~G.~M. Mitchell, and D.~J. Costello, ``Analyzing the
  finite-length performance of generalized {LDPC} codes,'' in \emph{2015 IEEE
  International Symposium on Information Theory (ISIT)}, 2015, pp. 2683--2687.

\bibitem{Liva10}
G.~Liva, W.~E. Ryan, and M.~Chiani, ``Quasi-cyclic generalized {LDPC} codes
  with low error floors,'' \emph{IEEE Transactions on Communications}, vol.~56,
  no.~1, pp. 49--57, 2008.

\bibitem{CHMRP19}
F.~Carpi, C.~Hager, M.~Martalo, R.~Raheli, and H.~D. Pfister, ``Reinforcement
  learning for channel coding: Learned bit-flipping decoding,'' \emph{Proc. of
  57th Allerton Conf. on Commun., Control and Computing}.

\bibitem{JSAIT}
S.~Habib, A.~Beemer, and Kliewer, ``Belief propagation decoding of short
  graph-based channel codes via reinforcement learning,'' \emph{IEEE Journal on
  Sel. Areas in Inf. Theory}, vol.~2, no.~2, pp. 627--640, 2021.

\bibitem{reldec}
S.~Habib, A.~Beemer, and J.~Kliewer, ``{RELDEC}: Reinforcement learning-based
  decoding of moderate length {LDPC} codes,'' [Online]. Available: arXiv.org,
  arXiv:2112.13934 [cs.IT], 2021.

\bibitem{Sutton18}
R.~S. Sutton and A.~G. Barto, \emph{Reinforcement Learning: An Introduction,
  2nd Edition}.\hskip 1em plus 0.5em minus 0.4em\relax The MIT Press Cambridge,
  2018.

\bibitem{Watkins89}
C.~J. C.~H. Watkins, ``Learning from delayed rewards,'' Ph.D. dissertation,
  King's College, 1989.

\bibitem{RU08}
T.~J. Richardson and R.~L. Urbanke, ``{M}odern {C}oding {T}heory,''
  \emph{Cambridge University Press}, 2008.

\end{thebibliography}

\end{document}